\documentclass[preprint,showpacs]{revtex4}
\usepackage{epsfig}
\usepackage{amsmath}
\usepackage{amssymb}

\begin{document}

\title{Branching ratio and CP violation of $B_{c}\to D K$  decays\\
in the perturbative QCD approach}

\author{Jun Zhang, Xian-Qiao Yu\footnote{yuxq@swu.edu.cn}}

\affiliation{
 {\it \small School of Physical Science and Technology, Southwest
 University, Chongqing 400715, China}}

\begin{abstract}
 In this paper, we calculate the
branching ratio and direct $CP$ asymmetry parameter of
$B_c^{\pm}\rightarrow D^{0}K^{\pm}$ in the framework of
perturbative QCD approach based on $k_T$ factorization. Besides
the usual factorizable diagrams, both non-factorizable and
annihilation type contributions are taken into account. We find
that (a) the branching ratio is at the order of $10^{-5}$; (b) the
tree annihilation diagrams and the penguin diagrams dominate the
total contribution; and (c) the direct CP asymmetry is about
$7\%$, which can be tested in the Large Hadron Collider beauty
experiments (LHC-b) at CERN.
\end{abstract}

 \maketitle

\section{Introduction}

  The rare B meson decays arouse more and more interest,
since it is a good place for testing the Standard Model (SM),
studying CP violation and looking for possible new physics beyond
the SM. In recent years, the theoretical studies of $B_{u,d}$
mesons have been done in the literature widely, which are strongly
supported by the running $B$ factories in KEK and Stanford Linear
Accelerator Center (SLAC). B physics studies are further supported
by the Large Hadron Collider beauty experiments (LHC-b), it is
estimated that about $5\times10^{10}$ $B_c$ mesons can be produced
per year at LHC\cite{LHC,AT}. So the studies of $B_{c}$ meson rare
decays are necessary in the next a few years.

The nonleptonic decays of the $B_c$ mesons have been studied in
previous
literature\cite{LHC,AT,Gouz,LL,GMM,Ivanov,Fajfer,EFG,Pakhomov,CMM,Verma,KKL,Saleev,CF,Fleischer,Munoz,Guo,Dai,Wei,JFLiu,DLY,Berezhnoi,Kiselev,Chang,XK,Masetti,DW,JFS,Cheng}
by employing Naive Factorization\cite{NF}, QCD
Factorization\cite{QCDF}, perturbative QCD approach
(PQCD)\cite{LY} and other approaches. The theoretical status of
the $B_c$ meson was reviewed in Ref.\cite{LHC}. In this paper, we
will study the $B_c\rightarrow DK$ decays in the perturbative QCD
approach. Our theoretical formulas for the decay $B_{c}\to D K$ in
PQCD framework are given in the next section. In section
\ref{sc:neval}, we give the numerical results of the branching
ratio and CP asymmetries of $B_{c}\to D K$  and the form factor of
$B_c \to D$ etc. At last, we give a short summary in section
\ref{summ}.

\begin{figure}[htb]
\vspace{0.5cm}
\begin{center}
\includegraphics[scale=0.75]{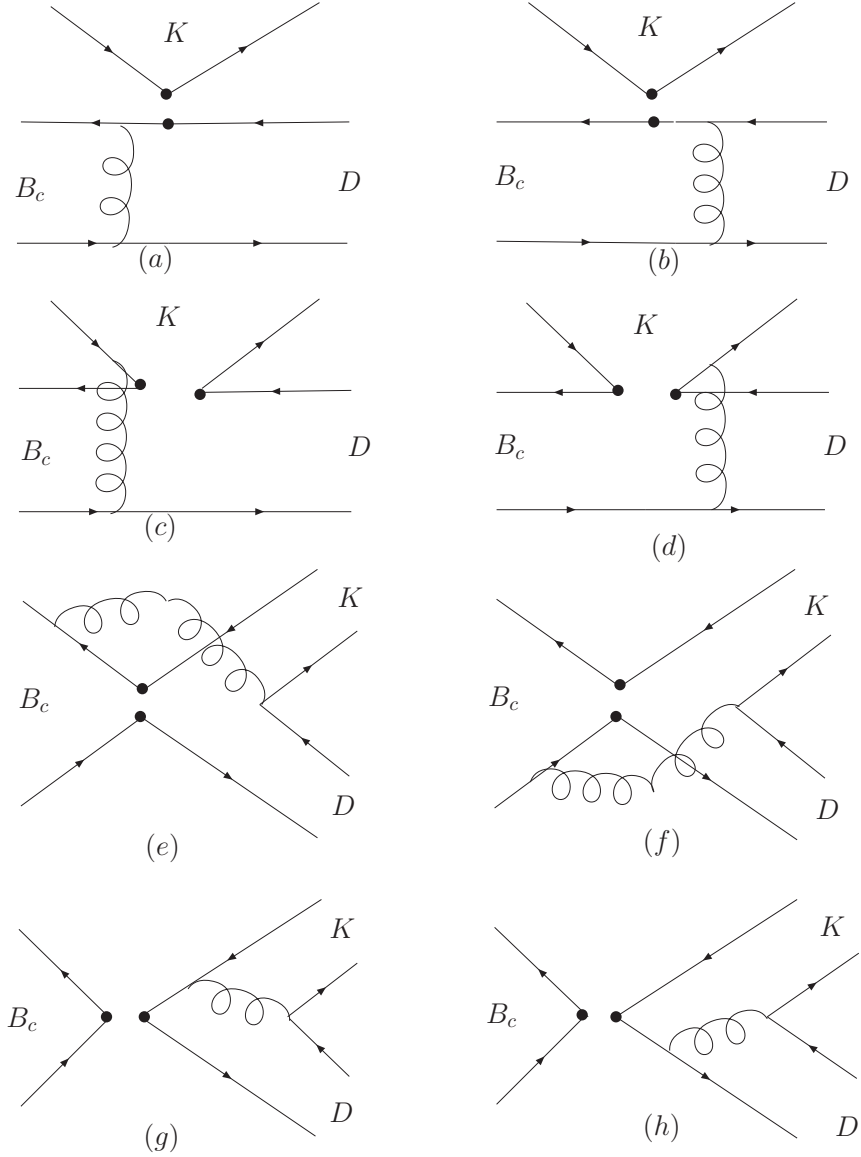}
\caption{The lowest order diagrams for $B_{c} \to D K$ decay.}
\label{figure:Fig1}
\end{center}
\end{figure}

\section{Perturbative calculations}\label{sc:fm}

For decay $B_{c}\to D K$, the related effective Hamiltonian is
given by \cite{BBL}
\begin{equation}
 H_\mathrm{eff} = \frac{G_F}{\sqrt{2}}\left\{\sum_{q=u,c} V_{qs}V_{qb}^* \left[
C_1(\mu) O_1^{q}(\mu) + C_2(\mu) O_2^{q}(\mu)
\right]-V_{tb}^*V_{ts}\sum_{i=3}^{10}C_i(\mu)
O_i(\mu)\right\},\label{hami}
\end{equation}
 where $C_{i}(\mu)(i=1,\cdots,10)$ are Wilson coefficients at the
  renormalization scale $\mu$ and $O_{i}(i=1,\cdots,10)$ are the four quark operators
\begin{equation}\begin{array}{ll}
  O_1^{q} = (\bar{b}_iq_j)_{V-A}(\bar{q}_js_i)_{V-A},  &
  O_2^{q} = (\bar{b}_iq_i)_{V-A} (\bar{q}_js_j)_{V-A},  \\
  O_3 = (\bar{b}_is_i)_{V-A}\sum_{q} (\bar{q}_jq_j)_{V-A},  &
  O_4 = (\bar{b}_is_j)_{V-A}\sum_{q} (\bar{q}_jq_i)_{V-A}, \\
  O_5 = (\bar{b}_is_i)_{V-A}\sum_{q} (\bar{q}_jq_j)_{V+A},  &
  O_6 = (\bar{b}_is_j)_{V-A} \sum_{q} (\bar{q}_jq_i)_{V+A}, \\
  O_7 = \frac{3}{2}(\bar{b}_is_i)_{V-A} \sum_{q}
   e_q(\bar{q}_jq_j)_{V+A},   &
   O_8 = \frac{3}{2}(\bar{b}_is_j)_{V-A}\sum_{q} e_q
  (\bar{q}_jq_i)_{V+A}, \\
  O_9 = \frac{3}{2}(\bar{b}_is_i)_{V-A}\sum_{q}
  e_q(\bar{q}_jq_j)_{V-A}, &
   O_{10} = \frac{3}{2}(\bar{b}_is_j)_{V-A}\sum_{q}
 e_q(\bar{q}_jq_i)_{V-A}. \label{eq:effectiv}
 \end{array}
\end{equation}
Here $i$ and $j$ are $SU(3)$ color indices; the sum over $q$ runs
over the quark fields that are active at the scale $\mu=O(m_{b})$,
i.e., $q\in \{u,d,s,c,b\}$. Operators $O_{1}, O_{2}$ come from
tree level interaction, while $O_{3}, O_{4}, O_{5}, O_{6}$ are
QCD-Penguins operators and $O_{7}, O_{8}, O_{9}, O_{10}$ come from
electroweak-penguins.

In PQCD approach, the decay amplitude can be written as:
\begin{equation}
 \mbox{Amplitude}
\sim \int\!\! d^4k_1 d^4k_2 d^4k_3\ \mathrm{Tr} \bigl[ C(t)
\Phi_{B_c}(k_1) \Phi_{D}(k_2) \Phi_K(k_3) H(k_1,k_2,k_3, t)
\bigr], \label{eq:convolution1}
\end{equation}
where $k_i$ are the momenta of light quarks included in each of
the mesons, and $\mathrm{Tr}$ denotes the trace over Dirac and
color indices. $C(t)$ is the Wilson coefficient results from the
radiative corrections at short distance. $\Phi_{M}$ is the wave
function which describes the hadronization of mesons. The wave
functions should be universal and channel independent, we can use
$\Phi_{M}$ which is determined by other ways. The hard part $H$
are channel dependent but fortunately perturbative calculable.

Working at the rest frame of $B_c$ meson, the momenta of the
$B_c$, $D$ and $K$ can be written in the light-cone coordinates as
:
\begin{equation}
       P_1 = \frac{M_{B_{c}}}{\sqrt{2}} (1,1,{\bf 0}_T),\ \ \  P_2 =
       \frac{M_{B_{c}}}{\sqrt{2}} (r^2,1,{\bf 0}_T), \ \ \ P_3 =
       \frac{M_{B_{c}}}{\sqrt{2}} (1-r^2,0,{\bf 0}_T) . \label{eq:momentun1}
\end{equation}
where $r=M_{D}/M_{B_{c}}$ and we neglect the kaon's mass $M_{K}$.
Denoting the light (anti-)quark momenta in $B_c$, $D$ and $K$ as
$k_1$, $k_2$ and $k_3$, respectively, we can choose:
\begin{equation}
k_1 = (x_1p_1^+, 0, {\bf k}_{1T}), ~~~k_2 = (0, x_2p_2^-, {\bf
k}_{2T}),~~~
 k_3 = (x_3p_3^+ , 0, {\bf k}_{3T}). \label{eq:momentun2}
\end{equation}
Then integration over $k_{1}^{-}, k_{2}^{+}$, and $k_{3}^{-}$ in
Eq.(\ref{eq:convolution1}) leads to
\begin{eqnarray}
 \mbox{Amplitude}
&&\sim  \int\!\! dx_1dx_2dx_3b_1db_1b_2db_2b_3db_3   \nonumber\\
 && \times \mathrm{Tr}
\bigl[ C(t) \Phi_{B_c}(x_1,b_1) \Phi_{D}(x_2,b_2) \Phi_K(x_3,b_3)
H(x_i,b_i,t) \bigr]e^{-S(t)}, \label{eq:convolution2}
\end{eqnarray}
where $b_i$ is the conjugate space coordinate of $k_{iT}$, and $t$
is the largest energy scale in $H$. The exponential Sudakov factor
$e^{-S(t)}$ comes from the higher order radiative corrections to
wave functions and hard amplitudes, it suppresses the soft
dynamics effectively\cite{Tseng} and thus makes a perturbative
calculation of the hard part $H$ reliable.

According to effective Hamiltonian (\ref{hami}), we draw the
lowest order diagrams of $B_{c}\rightarrow D K$ in Fig.
\ref{figure:Fig1}. The usual factorizable diagrams (a) and (b)
give the $B_c\to D$ form factor if take away the Wilson
coefficients. The operators $O_1, O_2, O_3, O_4, O_9$ and $O_{10}$
are $(V-A)(V-A)$ currents, and the sum of their contributions is
given by
\begin{eqnarray}
F_{e}[C]&=&\frac{8f_{B_{c}}}{\sqrt{2N_{c}}}\pi C_{F}
M_{B_{c}}^{2}\int_{0}^{1}dx_{2}\int_{0}^{\infty}b_{1}db_{1}b_{2}db_{2}\phi_{D}(x_{2},b_{2})
\nonumber\\
&&{}
\times\{[(r^{2}+2r-1)x_{2}+(2-r)r_{b}]\alpha_{s}(t_{a}^{1})h_{a}^{(1)}(x_{2},b_{1},b_{2})\exp[-S_{B}(t_{a}^{1}){}
\nonumber\\
&&{}-S_{D}(t_{a}^{1})]C(t_{a}^{1})+[r(2-r)]\alpha_{s}(t_{a}^{2})h_{a}^{(2)}(x_{2},b_{1},b_{2})\exp[-S_{B}(t_{a}^{2}){}
\nonumber\\
&&{}-S_{D}(t_{a}^{2})]C(t_{a}^{2})\},\label{eq:feT}
\end{eqnarray}
where $r_{b}=M_{b}/M_{B_{c}}$, $C_F=4/3$ is the
 group factor of the $SU(3)_{c}$ gauge group.
 The expressions of the meson distribution amplitudes
 $\phi_{M}$, the Sudakov factor $S_{X}(t_{i})(X=B_c, K, D)$,
 and the functions $h_a$ are given in the appendix. In above formula,
 the Wilson coefficients $C(t)$ of the corresponding operators
 are process dependent.

 The operator $O_5, O_6, O_7, O_8$ have the structure of $(V-A)(V+A)$,
  their amplitude is
\begin{eqnarray}
F_{e}^{P}[C]&=& \frac{16f_{B_{c}}}{\sqrt{2N_{c}}}r_{K}\pi C_{F}
M_{B_{c}}^{2}\int_0^1 dx_2\int_0^\infty b_{1}db_{1}
b_{2}db_{2}\phi_{D}(x_{2},b_{2}){}
\nonumber\\
&&{}\times\{[(2r^{2}-r)x_{2}+(2-r-r_{b}-2r^{2}+4r_{b}r)]\alpha_{s}(t_{a}^{1})h_{a}^{(1)}(x_{2},b_{1},b_{2}){}
\nonumber\\
&&{}\times\exp[-S_{B}(t_{a}^{1})-S_{D}(t_{a}^{1})]C(t_{a}^{1})+[3r-4r^{2}]\alpha_{s}(t_{a}^{2})h_{a}^{(2)}(x_{2},b_{1},b_{2}){}
\nonumber\\
&&{}\times\exp[-S_{B}(t_{a}^{2})-S_{D}(t_{a}^{2})]C(t_{a}^{2})\},\label{eq:feP}
\end{eqnarray}
where $r_K=M_{0K}/M_{B_{c}}=M_{K}^2/[M_{B_{c}}(M_s+M_u)]$. For the
non-factorizable diagrams (c) and (d), all three meson wave
functions are involved. Using $\delta$  function
$\delta(b_1-b_3)$, the integration of $b_1$ can be preformed
easily. For the $(V-A)(V-A)$ operators the result is:
\begin{eqnarray}
M_{e}[C]&=& \frac{16}{N_{c}}\pi C_{F} f_{B_{c}}
M_{B_{c}}^{2}\int_0^1 dx_{2}dx_{3}\int_0^\infty
b_{2}db_{2}b_{3}db_{3}\phi_{D}(x_{2},b_{2})\phi_{K}^{A}(x_{3}){}
\nonumber\\
&&{}\times\{[1-2r+(r-r^{2})x_{2}-(1-2r^{2})x_{3}]\alpha_{s}(t_{c}^{1})h_{c}^{(1)}(x_{2},x_{3},b_{2},b_{3}){}
\nonumber\\
&&{}\times
\exp[-S_{B}(t_{c}^{1})-S_{D}(t_{c}^{1})-S_{K}(t_{c}^{1})]C(t_{c}^{1})+[2r-1+(1-r-r^{2})x_{2}{}
\nonumber\\
&&{}-(1-2r^{2})x_{3}]\alpha_{s}(t_{c}^{2})h_{c}^{(2)}(x_{2},x_{3},b_{2},b_{3})\exp[-S_{B}(t_{c}^{2})-S_{D}(t_{c}^{2}){}
\nonumber\\
&&{}-S_{K}(t_{c}^{2})]C(t_{c}^{2})\},\label{eq:meT}
\end{eqnarray}
For the$(V-A)(V+A)$ operators, the formula is:
\begin{eqnarray}
M_{e}^{P}[C]&=& \frac{16}{N_{c}}\pi C_{F} f_{B_{c}} r_{K}
M_{B_{c}}^{2}\int_0^1 dx_{2}dx_{3}\int_0^\infty
b_{2}db_{2}b_{3}db_{3}\phi_{D}(x_{2},b_{2}){}
\nonumber\\
&&{}\times\{[(1+r-rx_{2}-(1+r)x_{3})\phi_{K}^{P}(x_{3})+(1-r+rx_{2}-(1+r)x_{3}){}
\nonumber\\
&&{}\times
\phi_{K}^{T}(x_{3})]\alpha_{s}(t_{c}^{1})h_{c}^{(1)}(x_{2},x_{3},b_{2},b_{3})\exp[-S_{B}(t_{c}^{1})-S_{D}(t_{c}^{1})-S_{K}(t_{c}^{1})]C(t_{c}^{1}){}
\nonumber\\
&&{}+[(rx_{2}-(1+r)x_{3})\phi_{K}^{P}(x_{3})+(-2r+rx_{2}+(1+r)x_{3})\phi_{K}^{T}(x_{3})]{}
\nonumber\\
&&{}\times\alpha_{s}(t_{c}^{2})h_{c}^{(2)}(x_{2},x_{3},b_{2},b_{3})\exp[-S_{B}(t_{c}^{2})-S_{D}(t_{c}^{2})-S_{K}(t_{c}^{2})]C(t_{c}^{2})\},\label{eq:meP}
\end{eqnarray}
Similar to (c),(d), the annihilation diagrams (e) and (f) also
involve all three meson wave functions. Here we have two kinds of
amplitudes, $M_a$ is the contribution containing the operator of
type $(V-A)(V-A)$, and $M_a^P$ is the contribution containing the
operator of type $(V-A)(V+A)$.
\begin{eqnarray}
M_{a}[C]&=&\frac{16}{N_{c}}\pi C_{F}
f_{B_{c}}M_{B_{c}}^{2}\int_{0}^{1}dx_{2}dx_{3}\int_{0}^{\infty}b_{1}db_{1}b_{2}db_{2}\phi_{D}(x_{2},b_{2}){}
\nonumber\\
&&{}\times
\{[(1-r-r_{b}-r^{2}x_{2}+(2r^{2}-1)x_{3}+r_{b}r^{2})\phi_{K}^{A}(x_{3})+(2-x_{2}-x_{3}-4r_{b}){}
\nonumber\\
&&{}\times
rr_{K}\phi_{K}^{P}(x_{3})+(x_{2}-x_{3})rr_{K}\phi_{K}^{T}(x_{3})]\alpha_{s}(t_{e}^{1})h_{e}^{(1)}(x_{2},x_{3},b_{1},b_{2})\exp[-S_{B}(t_{e}^{1}){}
\nonumber\\
&&{}-S_{D}(t_{e}^{1})-S_{K}(t_{e}^{1})]C(t_{e}^{1})+[(r+x_{2})\phi_{K}^{A}(x_{3})+(x_{2}+x_{3})rr_{K}\phi_{K}^{P}(x_{3}){}
\nonumber\\
&&{}+(x_{2}-x_{3})rr_{K}\phi_{K}^{T}(x_{3})]\alpha_{s}(t_{e}^{2})h_{e}^{(2)}(x_{2},x_{3},b_{1},b_{2})\exp[-S_{B}(t_{e}^{2})-S_{D}(t_{e}^{2}){}
\nonumber\\
&&{}-S_{K}(t_{e}^{1})]C(t_{e}^{2})\},\label{eq:maT}
\end{eqnarray}
\begin{eqnarray}
M_{a}^{P}[C]&=&\frac{16}{N_{c}}\pi C_{F}
f_{B_{c}}M_{B_{c}}^{2}\int_{0}^{1}dx_{2}dx_{3}\int_{0}^{\infty}b_{1}db_{1}b_{2}db_{2}\phi_{D}(x_{2},b_{2}){}
\nonumber\\
&&{}\times
\{[(-r+rx_{2}-r_{b}r)\phi_{K}^{A}(x_{3})+(1+r_{b}-r-x_{3})r_{K}\phi_{K}^{P}(x_{3})+(1+r_{b}{}
\nonumber\\
&&{}
-r-x_{3})r_{K}\phi_{K}^{T}(x_{3})]\alpha_{s}(t_{e}^{1})h_{e}^{(1)}(x_{2},x_{3},b_{1},b_{2})\exp[-S_{B}(t_{e}^{1})-S_{D}(t_{e}^{1}){}
\nonumber\\
&&{}-S_{K}(t_{e}^{1})]C(t_{e}^{1})+[(r^{2}-rx_{2})\phi_{K}^{A}(x_{3})+(x_{3}-2r)r_{K}\phi_{K}^{P}(x_{3}){}
\nonumber\\
&&{}+(x_{3}-2r)r_{K}\phi_{K}^{T}(x_{3})]\alpha_{s}(t_{e}^{2})h_{e}^{(2)}(x_{2},x_{3},b_{1},b_{2})\exp[-S_{B}(t_{e}^{2})-S_{D}(t_{e}^{2}){}
\nonumber\\
&&{}-S_{K}(t_{e}^{1})]C(t_{e}^{2})\},\label{eq:maP}
\end{eqnarray}

The amplitude for the factorizable annihilation diagrams (g) and
(h) result in $F_a$(for $(V-A)(V-A)$ type operators) and $F_a^P$
(for $(V-A)(V+A)$ type operators):
\begin{eqnarray}
F_{a}[C]&=& 16 \pi C_{F} M_{B_{c}}^{2}\int_0^1
dx_{2}dx_{3}\int_0^\infty
b_{2}db_{2}b_{3}db_{3}\phi_{D}(x_{2},b_{2}){}
\nonumber\\
&&{}\times\{[(r^{2}-1)x_{2}\phi_{K}^{A}(x_{3})-2rr_{K}(1+x_{2})\phi_{K}^{P}(x_{3})]\alpha_{s}(t_{g}^{1})h_{g}(x_{2},x_{3},b_{2},b_{3}){}
\nonumber\\
&&{}\times\exp[-S_{D}(t_{g}^{1})-S_{K}(t_{g}^{1})]C(t_{g}^{1})+[(-r^{2}+x_{3}-2r^{2}x_{3})\phi_{K}^{A}(x_{3}){}
\nonumber\\
&&{}+(r_{K}r+2r_{K}rx_{3})\phi_{K}^{P}(x_{3})+(-r_{K}r+2r_{K}rx_{3})\phi_{K}^{T}(x_{3})]{}
\nonumber\\
&&{}\times\alpha_{s}(t_{g}^{2})h_{g}(x_{3},x_{2},b_{3},b_{2})\exp[-S_{D}(t_{g}^{2})-S_{K}(t_{g}^{2})]C(t_{g}^{2})\},\label{eq:faT}
\end{eqnarray}
\begin{eqnarray}
F_{a}^{P}[C]&=& 32 \pi C_{F} M_{B_{c}}^{2}\int_0^1
dx_{2}dx_{3}\int_0^\infty
b_{2}db_{2}b_{3}db_{3}\phi_{D}(x_{2},b_{2}){}
\nonumber\\
&&{}\times\{[rx_{2}\phi_{K}^{A}(x_{3})+2r_{K}\phi_{K}^{P}(x_{3})]\alpha_{s}(t_{g}^{1})h_{g}(x_{2},x_{3},b_{2},b_{3})\exp[-S_{D}(t_{g}^{1}){}
\nonumber\\
&&{}
-S_{K}(t_{g}^{1})]C(t_{g}^{1})+[r\phi_{K}^{A}(x_{3})+r_{K}x_{3}\phi_{K}^{P}(x_{3})-r_{K}x_{3}\phi_{K}^{T}(x_{3})]{}
\nonumber\\
&&{}\times\alpha_{s}(t_{g}^{2})h_{g}(x_{3},x_{2},b_{3},b_{2})\exp[-S_{D}(t_{g}^{2})-S_{K}(t_{g}^{2})]C(t_{g}^{2})\},\label{eq:faP}
\end{eqnarray}

 From Equation (\ref{eq:feT})-(\ref{eq:faP}), the total decay amplitude
for $B_{c}^{+}\rightarrow D^{0}K^{+}$ can be written as
\begin{eqnarray}
A(B_{c}\rightarrow D^{0}K^{+}) &=&
f_{K}F_{e}[V_{us}V_{ub}^{*}(\frac{1}{3}C_{1}+C_{2})-V_{tb}^{*}V_{ts}(
  \frac{1}{3}C_{3}+C_{4}+\frac{1}{3}C_{9}+C_{10})]{}
  \nonumber\\
&&{} -f_{K}V_{tb}^{*}V_{ts}F_{e}^{P}[\frac{1}{3}C_{5}+C_{6}
  +\frac{1}{3}C_{7}+C_{8}]
  +M_{e}[V_{us}V_{ub}^{*}C_{1}-V_{tb}^{*}V_{ts}(C_{3}+C_{9})]{}
  \nonumber\\
&&{}-V_{tb}^{*}V_{ts}M_e^{P}(C_{5}+C_{7})+M_a[V_{cs}V_{cb}^{*}C_{1}-V_{tb}^{*}V_{ts}(C_{3}
  +C_{9})]{}
  \nonumber\\
  &&{}-V_{tb}^{*}V_{ts}M_a^P(C_{5}+C_{7})+f_{B_{c}}F_{a}[V_{cs}V_{cb}^{*}(\frac{1}{3}C_{1}+C_{2})-V_{tb}^{*}V_{ts}(
  \frac{1}{3}C_{3}+C_{4}{}
  \nonumber\\
  &&{}+\frac{1}{3}C_{9}+C_{10})]-f_{B_{c}}V_{tb}^{*}V_{ts}F_a^{P}[\frac{1}{3}C_{5}+C_{6}+\frac{1}{3}C_{7}
   +C_{8}],\label{eq:da}
\end{eqnarray}
and the decay width is expressed as
\begin{equation}
 \Gamma(B_c^+ \to D^0 K^+) = \frac{G_F^2 M_{B_{c}}^3}{128\pi}(1-r^2)
|A(B_{c}^+\to D^{0}K^{+})|^2. \label{eq:dw}
\end{equation}
The decay amplitude of the charge conjugate channel $B_{c}^{-}\to
\bar{D}^{0}
 K^-$ can be obtained by replacing $V_{qs}V_{qb}^{*}$ to
 $V_{qs}^{*}V_{qb}$ and $V_{tb}^{*}V_{ts}$ to $V_{tb}V_{ts}^{*}$
  in Eq.(\ref{eq:da}).

Using the unitary condition of the CKM matrix elements
$V_{ub}^{*}V_{us}+ V_{cb}^{*}V_{cs}+V_{tb}^{*}V_{ts}=0$, the decay
amplitude of $B_{c}^{+}\rightarrow D^{0}K^{+}$ in Eq.(\ref{eq:da})
can be parameterized as
 \begin{eqnarray}
A &=& V_{ub}^*V_{us}T_{u}+V_{cb}^*V_{cs}T_{c}-V_{tb}^*V_{ts}P{}
\nonumber\\
 &=&V_{ub}^*V_{us}(T_{u}+P)[1+z
e^{i(\delta-\gamma)}],\label{a1}
\end{eqnarray}
where
$z=|\frac{V_{cb}^*V_{cs}}{V_{ub}^*V_{us}}||\frac{T_c+P}{T_u+p}|$,
$V_{ub}\simeq |V_{ub}|e^{-i\gamma}$ and
$\delta=\arg(\frac{T_c+P}{T_u+P})$. $z$ and $\delta$ can be
calculated from PQCD.

Similarly, the decay amplitude for $B_{c}^{-}\rightarrow
\bar{D}^{0}K^{-}$ can be parameterized as
 \begin{eqnarray}
  \bar{A} &=&
V_{ub}V_{us}^{*}T_{u}+V_{cb}V_{cs}^{*}T_{c}-V_{tb}V_{ts}^{*}P{}
\nonumber\\
 &=&V_{ub}V_{us}^{*}(T_{u}+P)[1+z
e^{i(\delta+\gamma)}],\label{a2}
\end{eqnarray}
and the averaged decay width for $B_{c}^{+}(B_{c}^{-})\rightarrow
D^{0} K^{\pm}$ reads
\begin{eqnarray}
 \Gamma(B_{c}^{+}(B_{c}^{-}) \to D^{0} K^{\pm}) &=& \frac{G_F^2 M_{B_{c}}^3}{128\pi}(1-r^2)
(|A|^2/2+|\bar{A}|^2/2)\hspace*{1cm}  \nonumber \\
&=&\frac{G_F^2
M_{B_{c}}^3}{128\pi}(1-r^2)|V_{ub}^*V_{us}(T_{u}+P)|^{2}[1+2z\cos\gamma
\cos\delta+z^{2}], \label{eq:width3}
\end{eqnarray}
which is a function of CKM angle $\gamma$.

\section{Numerical evaluation}\label{sc:neval}

The following parameters have been used in our numerical
calculation  \cite{PDG,TWQCD,nari,Chang,CDLu}:
\begin{gather}
\nonumber  M_{B_c} = 6.286 \mbox{GeV}, M_{b} = 4.8 \mbox{GeV},
M_{D} = 1.685 \mbox{GeV}, M_{0K} = 1.6 \mbox{GeV},
 \\ \nonumber \omega_{D}=0.2\mbox{GeV}, f_{B_c} = 489 \mbox{MeV},
 f_{K} = 160 \mbox{MeV}, f_{D} = 240 \mbox{MeV},
 \\ a_{D}=0.3,  \tau_{B_c}=0.46\times 10^{-12}\mbox{s},
  |V_{ub}^{*}V_{us}|=0.0009, |V_{cb}^{*}V_{cs}|=0.0398.
\label{eq:shapewv}
\end{gather}
\begin{figure}[htbp!]
\begin{center}
\includegraphics[width=0.7 \textwidth]{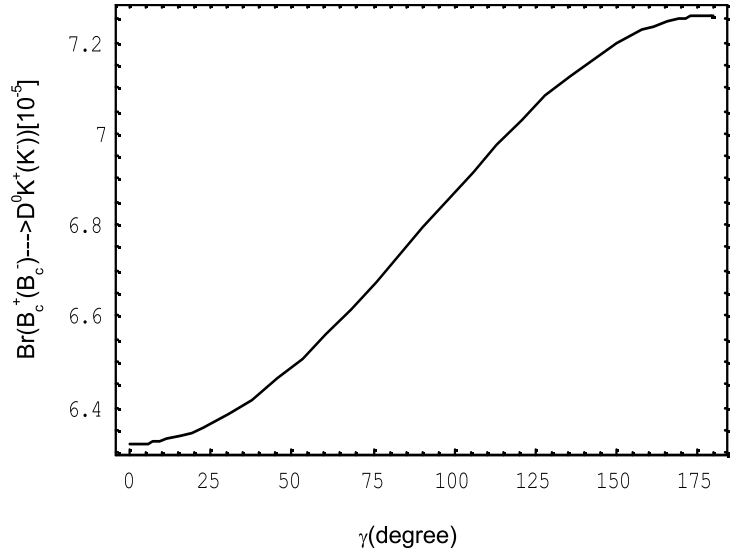}
\vspace{-0.5cm} \caption{The averaged branching ratio  of
    $B_{c}^{+}(B_{c}^{-}) \to D^{0} K^{\pm}$ decay as a function of CKM angle
    $\gamma$.} \label{figure:Fig2}
\end{center}
\end{figure}
We leave the CKM phase angle $\gamma$ as a free parameter to explore
the branching ratio and CP asymmetry parameter dependence on it. The
averaged branching ratio of the decay $B_{c}^{+}(B_{c}^{-}) \to
D^{0} K^{\pm}$ with respect to the parameter $\gamma$ is shown in
Fig. \ref{figure:Fig2}. Since the CKM angle $\gamma$ is constrained
as $\gamma=(63^{+15}_{-12})^{\circ}$\cite{PDG}, we can arrive from
Fig. \ref{figure:Fig2}:
\begin{equation}
    6.5\times10^{-5}<Br(B_{c}^{+}(B_{c}^{-}) \to D^{0} K^{\pm})<6.7\times10^{-5},  \hspace{5mm}\text{for} 50^{\circ}<\gamma<80^{\circ}.
\end{equation}

Our numerical analysis show that
$|V_{cb}^*V_{cs}T_c/V_{ub}^*V_{us}T_u|=10$ and
$|V_{tb}^*V_{ts}P/V_{ub}^*V_{us}T_u|=7.4$ which mean the tree
level contributions from annihilation topology and the penguin
contributions dominate in this decay and the branching ratio is
not sensitive to $\gamma$. In our calculation, the only input
parameters are wave functions, which represent the
non-perturbative contributions. In all the three meson wave
functions, the main uncertainty come from the value of $\omega_D$
in $D$ meson wave function(see appendix). We investigate the
branching ratio dependence on the value of $\omega_D$ in Table
\ref{tab1}. By changing the value of $\omega_D$ from
$0.2GeV$\cite{CDLu} to $0.4GeV$, we find the branching ratio of
$B_{c}^{+}(B_{c}^{-}) \to D^{0} K^{\pm}$ change little as shown in
Table \ref{tab1}.

\begin{table}[htb]
\begin{center}
\begin{tabular}[t]{r|c|c|c}
\hline \hline
      & $\omega_{D}=0.2GeV$ &  $\omega_{D}=0.3GeV$ & $\omega_{D}=0.4GeV$ \\
  \hline
  $B_{c}^{+}\rightarrow D^{0} K^{+}$
  & $6.6$  & $7.0$  & $5.6$\\
\hline \hline
\end{tabular}
\end{center}
\caption{Branch ratios in the unit $10^{-5}$ using $\gamma=60^\circ$
for different $\omega_{D}$} \label{tab1}
\end{table}

The diagrams (a) and (b) in Fig. \ref{figure:Fig1} give the
 contribution for $B_{c}\rightarrow D$ transition form factor
  $F^{B_{c}\to D}(q^{2}=M_{K}^2\simeq 0)$, where $q=P_{1}-P_{2}$ is the momentum
   transfer. The sum of their amplitudes have been given by Eq.~(\ref{eq:feT}),
   so we can use PQCD approach to compute this form factor.
  Our result is $F^{B_{c}\to D}(0)=0.24$, if $\omega_D=0.2GeV$; and
  $F^{B_{c}\to D}(0)=0.21$, if $\omega_D=0.45GeV$. We can see that
   this form factor is not
   sensitive to the $D$ meson wave function. In the literature,
   there already exist a lot of studies on $B_{c}\rightarrow D$ transition form
   factor\cite{DW,CNP,NW,IKS,KKL,EFG,ZH,DSV,WSL}, we show their
   results in Table \ref{tab2}. From which we find that there
   are large differences in these results(including ours) and
   eventually this form factor can be extracted from
  semi-leptonic experiments $B_{c}\rightarrow D l\nu_{l}$ in the
  future LHC experiments.

 \begin{table}[htb]
\begin{center}
\begin{tabular}[t]{r|c|c|c|c|c|c|c|c|c}
\hline \hline
      & DW\cite{DW}\footnote{We quote the result with $\omega=0.7
\mbox{GeV}$.} &  CNP\cite{CNP} & NW\cite{NW} & IKS\cite{IKS} &
KKL\cite{KKL}\footnote{The nonbracketed(bracketed) result is
evaluated in sum rules(potential model).} &
      EFG\cite{EFG}
      & ZH\cite{ZH} & DSV\cite{DSV} & WSL\cite{WSL} \\
  \hline
  $F^{B_{c}\to D}(0)$
  & $0.255$  & $0.13$  & $0.1446$ & $0.69$ & $0.32[0.29]$ & $0.14$ & $0.35$ & $0.075$ & $0.16$  \\
\hline \hline
\end{tabular}
\end{center}
\caption{$B_{c}\rightarrow D$ transition form factor at $q^{2}=0$
evaluated in the literature.} \label{tab2}
\end{table}

\begin{figure}[htbp!]
\begin{center}
\includegraphics[width=0.7 \textwidth]{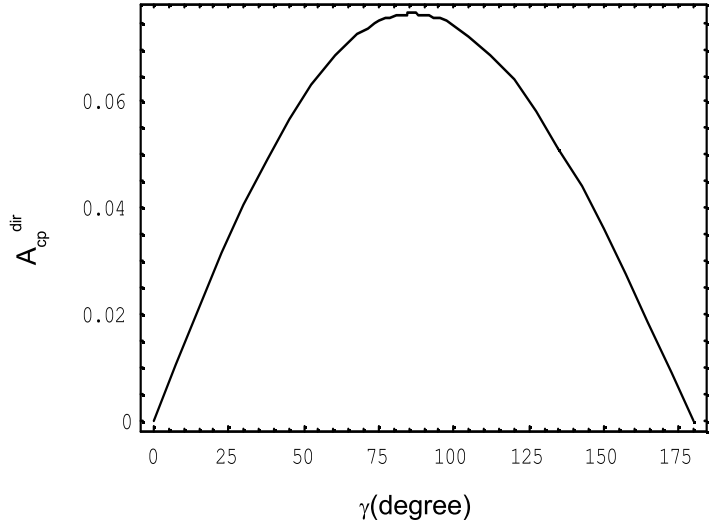}
\vspace{-0.5cm} \caption{Direct CP violation parameters of
$B_{c}^{\pm}\to D^{0}K^{\pm}$ decay as a function of CKM angle
$\gamma$.}\label{figure:Fig3}
\end{center}
\end{figure}

  The direct CP violation $A_{CP}^{dir}$ is defined as
\begin{eqnarray}
A_{CP}^{dir}&=&\frac{\Gamma(B_{c}^{+}\rightarrow
D^{0}K^{+})-\Gamma(B_{c}^{-}\rightarrow
\bar{D}^{0}K^{-})}{\Gamma(B_{c}^{+}\rightarrow
D^{0}K^{+})+\Gamma(B_{c}^{-}\rightarrow \bar{D}^{0}K^{-})}{}
\nonumber\\
&=&\frac{2z\sin\gamma\sin\delta}{1+2z\cos\gamma\cos\delta+z^{2}}.
 \label{dcpv}
 \end{eqnarray}
  Using Eq.(\ref{a1}) and (\ref{a2}), we compute the parameter
 $A_{CP}^{dir}$. The direct CP asymmetry  $A_{CP}^{dir}$ has a strong dependence
 on the CKM angle $\gamma$, as can be seen easily from Eq.(\ref{dcpv}) and Fig. \ref{figure:Fig3}.
 From this figure one can see that the direct CP asymmetry
 at $6\%-7.6\%$ for $50^{\circ}<\gamma<80^{\circ}$. The small direct CP
 asymmetry is also a result of small tree level contribution from emission topology.
\par
\section{Summary} \label{summ}
 In this work, we study the branching ratio and CP asymmetry of
  the decays $B_{c}^{\pm}\to D^{0}K^{\pm}$
  in PQCD approach. It is found that the branching
  ratio of $B_{c}^{\pm}\to D^{0}K^{\pm}$ is at the order of $10^{-5}$. We also predict $CP$ asymmetries in the process, which
 may be measured in the LHC-b experiments.

\section*{Acknowledgments}
 This work is supported by the National Natural Science Foundation of
 China under Grant Nos. 10847157, 10575083. The authors would like to thank M-Z Zhou for useful discussions.


\begin{appendix} \label{appendix}

\section{formulas for the calculations used in the text}

 In the appendix we present the explicit expressions of the
 formulas used in section II. First, we give the expressions of the
 meson distribution amplitudes
 $\phi_{M}$. For $B_{c}$ meson wave function, we use the function
 as\cite{Cheng}:
\begin{equation}
\Phi_{B_c}(x) =\frac{f_{B_c}}{4N_c}(\not \!
p_{B_c}+M_{B_c})\gamma_5\delta(x-M_c/M_{B_c}) .\label{waveb}
\end{equation}

For $D$ meson wave function, we use the function
 as\cite{Kurimoto}:
\begin{equation}
\phi_{D}(x,b)
=\frac{3}{\sqrt{2N_c}}f_{D}x(1-x)\{1+a_{D}(1-2x)\}\exp[-\frac{1}{2}(\omega_{D}b)^2],
\label{wavedd}
\end{equation}
We set $a_{D}=0.3$ and $\omega_{D}=0.2GeV$\cite{CDLu} in our
numerical calculation.

We use the distribution amplitude $\phi^{A,P,T}_{K}$ of the K
meson from
 Ref.  \cite{atm}:
 \begin{eqnarray}
\nonumber \phi_{K}^A(x) &=& \frac{6f_{K}}{2\sqrt{2
N_c}}x(1-x)[1+0.15t+0.405(5t^2-1)], \\
\nonumber \phi_{K}^P(x) &=& \frac{f_{K}}{2\sqrt{2 N_c}} [1
+0.106(3t^2-1)-0.148(3-30t^2+35t^4)/8], \\
\phi_{K}^T(x) &=& \frac{f_{K}}{2\sqrt{2 N_c}}t[1+0.1581(5t^2-3)],
\end{eqnarray}
where $t=1-2x$. whose coefficients correspond to $m_{0K} = 1.6
\mbox{GeV}$.

 $S_{B_c}$, $S_{D}$, $S_{K}$ used in the decay amplitudes
 are defined as
\begin{gather}
 S_{B_c}(t)=s(x_1P_1^+,b_1)+2\int_{1/b_1}^t\!\!\!\frac{d\bar\mu}{\bar\mu}
 \gamma(\alpha_s(\bar\mu)),\hspace*{3.5cm}\\
 S_{D}(t)=s(x_2P_2^-,b_2)+
 2\int_{1/b_2}^t\!\!\!\frac{d\bar\mu}{\bar\mu}\gamma(\alpha_s(\bar\mu)),\hspace*{3.5cm}\\
 S_{K}(t)=s(x_3P_3^+,b_3)+s((1-x_3)P_3^+,b_3)+
 2\int_{1/b_3}^t\!\!\!\frac{d\bar\mu}{\bar\mu}\gamma(\alpha_s(\bar\mu)),
\end{gather}
where the so called Sudakov factor $s(Q,b)$  resulting from the
resummation of double logarithms is given as \cite{LU,LM}
\begin{equation}
s(Q,b)=\int_{1/b}^Q\!\!\! \frac{d\mu}{\mu}\Bigl[
\ln\left(\frac{Q}{\mu}\right)A(\alpha(\bar\mu))+B(\alpha_s(\bar\mu))
\Bigr] \label{su1}
\end{equation}
with
\begin{gather}
A=C_F\frac{\alpha_s}{\pi}+\left[\frac{67}{9}-\frac{\pi^2}{3}-\frac{10}{27}n_{f}+
\frac{2}{3}\beta_0\ln\left(\frac{e^{\gamma_E}}{2}\right)\right]
 \left(\frac{\alpha_s}{\pi}\right)^2 ,\\
B=\frac{2}{3}\frac{\alpha_s}{\pi}\ln\left(\frac{e^{2\gamma_{E}-1}}{2}\right).\hspace{6cm}
\end{gather}
Here $\gamma_E=0.57722\cdots$ is the Euler constant, $n_{f}$ is
the active quark flavor number.

The functions $h_{i}(i=a,c.e.g)$   come from the Fourier
transformation of propagators of virtual quark and gluon in the
hard part calculations. They are given as
\begin{align}
& h_{a}^{(1)}(x_2,b_1,b_2) = S_{t}(x_2)K_{0}(M_{B}\sqrt{(1-x_2)(r-r^2)}b_{1})\nonumber \\
&
  \times
 [\theta(b_2-b_1)\theta(r_{b}^{2}-(1-r^{2})x_{2})I_0(M_B\sqrt{r_{b}^2-(1-r^2)x_2}b_1)\nonumber \\
 &
  \times
    \mathrm{K}_0(M_B\sqrt{r_{b}^2-(1-r^2)x_2}b_2)
 +(b_1\leftrightarrow b_2)],
\end{align}

\begin{align}
& h_{a}^{(2)}(x_2,b_1,b_2) = S_{t}(r)K_{0}(M_{B}\sqrt{(1-x_2)(r-r^2)}b_{2})\nonumber \\
&
  \times \bigl[\theta(b_2-b_1)I_{0}(M_{B}\sqrt{r-r^2}b_1)K_{0}(M_{B}\sqrt{r-r^2}b_2)+(b_1
\leftrightarrow b_2)\bigr],
 \label{eq:propagator1}
\end{align}

\begin{align}
& h_{c}^{(j)}(x_2,x_3,b_2,b_3) = \nonumber \\
&
\biggl\{\theta(b_2-b_3)I_{0}(M_B\sqrt{(1-x_2)(r-r^2)}b_3)K_{0}(M_B\sqrt{(1-x_2)(r-r^2)}b_2)
\nonumber \\
& \qquad\qquad\qquad\qquad + (b_2\leftrightarrow b_3) \biggr\}
 \times\left(
\begin{matrix}
 \mathrm{K}_0(M_B F_{(j)} b_3), & \text{for}\quad F^2_{(j)}>0 \\
 \frac{\pi i}{2} \mathrm{H}_0^{(1)}(M_B\sqrt{|F^2_{(j)}|}\ b_3), &
 \text{for}\quad F^2_{(j)}<0
\end{matrix}\right),
\label{eq:propagator2}
\end{align}
where $\mathrm{H}_0^{(1)}(z) = \mathrm{J}_0(z) + i\,
\mathrm{Y}_0(z)$, and $F_{(j)}$'s are defined by
\begin{equation}
 F^2_{(1)} = (1-r-x_3+r^2x_3)(x_2-1),\
F^2_{(2)} = (1-x_2)(r-r^2-x_3+r^2x_3);
\end{equation}
\begin{align}
& h^{(j)}_e(x_2,x_3,b_1,b_2) = \nonumber \\
& \biggl\{\theta(b_2-b_1) \frac{\pi i}{2}
\mathrm{H}_0^{(1)}(M_B\sqrt{(1-r^2)x_2x_3}\, b_2)
 \mathrm{J}_0(M_B\sqrt{(1-r^2)x_2x_3}\, b_1)
\nonumber \\
& \qquad\qquad\qquad\qquad + (b_1 \leftrightarrow b_2) \biggr\}
 \times\left(
\begin{matrix}
 \mathrm{K}_0(M_B F_{e(j)} b_1), & \text{for}\quad F^2_{e(j)}>0 \\
 \frac{\pi i}{2} \mathrm{H}_0^{(1)}(M_B\sqrt{|F^2_{e(j)}|}\ b_1), &
 \text{for}\quad F^2_{e(j)}<0
\end{matrix}\right),
\label{eq:propagator3}
\end{align}
where $F_{e(j)}$'s are defined by
\begin{equation}
 F^2_{e(1)} =r_{b}^2-(1-x_2)(1-r-x_3+r^2x_3),\
F^2_{e(2)} = r^2-x_2(x_3-r-r^2x_3);
\end{equation}
\begin{align}
& h_{g}(x_2,x_3,b_2,b_3) = S_{t}(x_2)\frac{\pi i}{2}H_{0}^{(1)}
(M_{B}\sqrt{(1-r^2)x_2x_3}b_{3})\nonumber \\
&\times
\bigl[\theta(b_3-b_2)J_{0}(M_{B}\sqrt{(1-r^2)x_2}b_2)\frac{\pi i}{2}
  H_{0}^{(1)}(M_{B}\sqrt{(1-r^2)x_2}b_3)+(b_2
\leftrightarrow b_3)\bigr].
 \label{eq:propagator4}
\end{align}

 We adopt the parametrization for $S_{t}(x)$ contributing to the
 factorizable diagrams  \cite{KLS},
 \begin{align}
S_{t}(x)=\frac{2^{1+2c}\Gamma(3/2+c)}{\sqrt{\pi}\Gamma(1+c)}[x(1-x)]^{c},
\hspace{0.5cm}c=0.3.
\end{align}
The hard scale $t_{i}'s$ in Eq.(\ref{eq:feT})-(\ref{eq:faP}) are
chosen as
 \begin{eqnarray}
\nonumber t_a^{1} &=& \mathrm{max}(M_B \sqrt{(1-x_{2})(r-r^2)},M_B
\sqrt{|r_{b}^{2}-(1-r^2)x_2|},
 1/b_1,1/b_2), \\
\nonumber t_a^{2} &=& \mathrm{max}(M_B \sqrt{r-r^2},
 1/b_1,1/b_2), \\
\nonumber t_c^{1} &=& \mathrm{max}(M_B
\sqrt{|F^2_{(1)}|},M_B\sqrt{(r-r^2)(1-x_2)},
 1/b_2,1/b_3), \\
 \nonumber t_c^{2} &=& \mathrm{max}(M_B
\sqrt{|F^2_{(2)}|},M_B\sqrt{(r-r^2)(1-x_2)},
 1/b_2,1/b_3), \\
 \nonumber t_e^{1} &=& \mathrm{max}(M_B
\sqrt{|F^2_{e(1)}|},M_B\sqrt{(1-r^2)x_{2}x_3},
 1/b_1,1/b_2), \\
 \nonumber t_e^{2} &=& \mathrm{max}(M_B
\sqrt{|F^2_{e(2)}|},M_B\sqrt{(1-r^2)x_{2}x_3},
 1/b_1,1/b_2), \\
 \nonumber t_g^{1} &=& \mathrm{max}(M_B \sqrt{(1-r^2)x_{2}},
 1/b_2,1/b_3), \\
 t_g^{2} &=& \mathrm{max}(M_B \sqrt{(1-r^2)x_{3}},
 1/b_2,1/b_3).
\end{eqnarray}
 They are given as the maximum energy scale appearing in each diagram
 to kill the large logarithmic radiative corrections.
\end{appendix}


\end{document}